\documentclass[10pt,journal,final]{IEEEtran}
\usepackage{ifpdf}

\ifCLASSOPTIONcompsoc
  \usepackage[numbers,sort&compress]{natbib}
\else
  \usepackage{cite}
\fi

\ifCLASSINFOpdf
  \usepackage[pdftex]{graphicx}
  \DeclareGraphicsExtensions{.pdf,.jpeg,.png}
\else
  \usepackage[dvips]{graphicx}
\fi

\usepackage[cmex10]{amsmath}
\usepackage{algorithmic}
\usepackage{array}
\usepackage{url}
\usepackage{multirow}
\usepackage{algorithm}
\usepackage{algorithmic}
\usepackage{amsmath}
\usepackage{amssymb}
\usepackage{balance}
\usepackage{color, soul}
\usepackage{bm}

\ifCLASSINFOpdf
  \usepackage[pdftex]{thumbpdf}
\else
\usepackage[dvips]{thumbpdf}
\fi

\usepackage{threeparttable}
\usepackage{multirow}
\usepackage{hyperref}
\usepackage{textcomp}

\hypersetup{colorlinks=true, linkcolor=blue, anchorcolor=blue, citecolor=blue, filecolor=blue, pagecolor=blue, urlcolor=blue}

\begin{document}
\title{Higher-order Fuzzy Membership in Motif Modularity Optimization}

\author{Jing~Xiao, Ya-Wei Wei, and Xiao-Ke Xu,~\IEEEmembership{Member,~IEEE}
\thanks{Manuscript received xxxx. This work was supported by the Beijing Natural Science Foundation under Grant 4242040, National Natural Science Foundation of China under Grants 62173065 and 61603073, Fundamental Research Funds for the Central Universities under Grant 124330008, and Basic Research Projects of Liaoning Provincial Department of Education under Grant LJKMZ20220399. (Corresponding author: Xiao-Ke Xu.)}
\IEEEcompsocitemizethanks{\IEEEcompsocthanksitem Jing~Xiao is with the College of Big Data and Internet, Shenzhen Technology University, Shenzhen 518118, China, and also with the College of Information and Communication Engineering, Dalian Minzu University, Dalian 116600, China. Ya-Wei Wei is with the College of Information and Communication Engineering, Dalian Minzu University, Dalian 116600, China. Xiao-Ke Xu is with the School of Journalism and Communication, Beijing Normal University, Beijing 100875, China (e-mail: xuxiaoke@foxmail.com).}}

\markboth{}%
{Shell \MakeLowercase{\textit{et al.}}: Bare Demo of IEEEtran.cls for Computer Society Journals}

\IEEEtitleabstractindextext{
\begin{abstract}
Higher-order community detection (HCD) reveals both mesoscale structures and functional characteristics of real-life networks. Although many methods have been developed from diverse perspectives, to our knowledge, none can provide fine-grained higher-order fuzzy community information. This study presents a novel concept of higher-order fuzzy memberships that quantify the membership grades of motifs to crisp higher-order communities, thereby revealing the partial community affiliations. Furthermore, we employ higher-order fuzzy memberships to enhance HCD via a general framework called fuzzy memberships assisted motif-based evolutionary modularity (FMMEM). In FFMEM, on the one hand, a fuzzy membership-based neighbor community modification (FM-NCM) strategy is designed to correct misassigned bridge nodes, thereby improving partition quality. On the other hand, a fuzzy membership-based local community merging (FM-LCM) strategy is also proposed to combine excessively fragmented communities for enhancing local search ability. Experimental results indicate that the FMMEM framework outperforms state-of-the-art methods in both synthetic and real-world datasets, particularly in the networks with ambiguous and complex structures. 
\end{abstract}

\begin{IEEEkeywords}
Higher-order community detection, network motif, fuzzy membership, modularity optimization.
\end{IEEEkeywords}}

\maketitle
\IEEEdisplaynontitleabstractindextext
\IEEEpeerreviewmaketitle

\section{Introduction}

\IEEEPARstart{C}{ommunity} structure, ubiquitous in social \cite{girvan2002community}, biological \cite{Zhu2021Nodal} and engineering networks \cite{fortunato2016community}, is one of the significant characteristics of complex networks. Community detection serves as a potent analytical tool for exploring both explicit topologies and implicit functional characteristics of real-life networks \cite{javed2018community, Mittal2021Classification}. Despite the development of numerous community detection methods grounded in diverse knowledge domains \cite{Baraa2021A, Zhang2020A, Chao2021A}, most of them mainly focus on discovering communities using lower-order connectivity patterns at the level of nodes and edges \cite{Benson2016Higher}. Consequently, the higher-order connectivity patterns widespread in real-life networks are largely overlooked \cite{Huang2021HM-Modularity}.

Recently, higher-order community detection (HCD) has attracted significant attention due to its unique ability to leverage higher-order connectivity patterns, namely, network motifs \cite{Milo2002Network}, to reveal higher-order organizational structures of real-life networks \cite{Benson2016Higher}. Specifically, network motifs are essentially small and frequently recurring subgraphs, generally considered as fundamental building blocks and functional units of real-life networks\cite{Huang2021HM-Modularity, Lotito2022HigherorderMA, Huang2020Efficient}. Compared to traditional lower-order connectivity patterns (e.g., edges), network motifs not only can better signature community structures but also typically reflect specific functions, such as traffic flow in transportation networks and information processing in neuron networks \cite{Lotito2022HigherorderMA}. Essentially, HCD based on a specific motif aims to partition a network (graph) into groups of nodes where intra-connected motif instances are dense, and inter-connected motif instances are sparse \cite{Zhang2020A,Pizzuti2017An}. Typically, HCD can usually reveal more precise and meaningful communities, offering deeper insights into network structures and functional characteristics \cite{Huang2020Efficient}.

Many HCD methods have been developed from diverse perspectives, including motif-based spectral clustering \cite{Benson2016Higher, Tsourakakis2017Scalable, Lim2016Motif}, motif-based local expansion \cite{ma2019local, Yin2017Local}, motif-based label propagation \cite{li2020community, Li2021MotifbasedEL}, and motif-based connection enhancement \cite{li2019EdMot}, etc. Particularly, motif-based optimization methods garnered significant attention \cite{Huang2021HM-Modularity}, due to the advantages of approximating the global optimum, stability, and few parameters. Essentially, they model HCD as a global optimization problem with a specific motif-based objective function, such as motif conductance \cite{Benson2016Higher}. Despite the effectiveness of motif-based optimization methods in identifying higher-order communities, they encounter two main challenges: none can provide fine-grained higher-order fuzzy community information, and the limitation in quality of identified higher-order communities due to the absence of detailed network topological features, particularly in ambiguous structures.

For one thing, to our knowledge, no methods can provide detailed higher-order fuzzy community information. Although lower-order fuzzy memberships have been developed to provide more fine-grained membership grades for nodes valued continuously between $[0, 1]$, they primarily focus on lower-order crisp communities formed at the edge level \cite{Yazdanparast2021Soft,su2015quadratic,Xiao2021Fuzzy}, which are not suitable for higher-order community structures. Furthermore, the higher-order communities identified by previous HCD methods are generally characterized by coarse-grained crisp community structures, where each motif can only full belong to one community, i.e., the membership grade is binary (0 or 1) \cite{Benson2016Higher}. However, in real-world networks, a motif may belong to multiple communities and has partial membership to each. For example, in a scientific collaboration network, a team of several scientists (treated as a motif) may be involved to varying degrees in multiple academic organizations (treated as communities). Consequently, the partial belongingness of a motif across multiple communities is largely overlooked.

\begin{figure}[ht]
\centering
\includegraphics[width=0.30\textwidth]{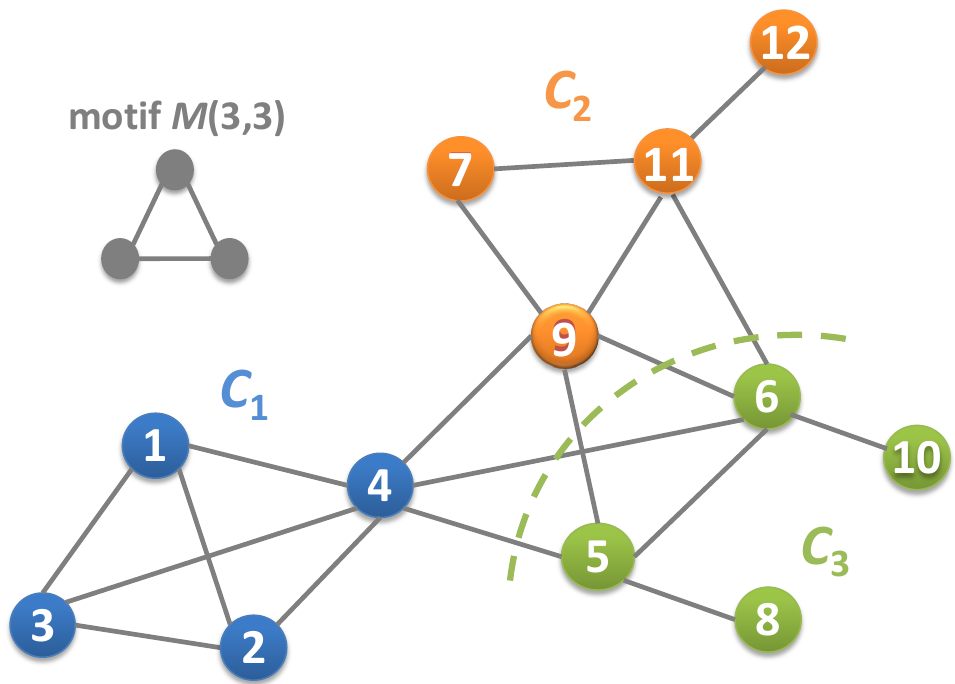}
\caption{A higher-order partition of a simple network identified using typical HCD methods (e.g., MELPA \cite{Li2021MotifbasedEL} and EdMot \cite{li2019EdMot}) with a triangular motif $M\left (3,3\right )$. The similar motif connections from bridge node 4 to community $C_{1}$ and others makes node 4 easily to be misassigned. Additionally, $C_{3}$ appears overly fragmented because it has denser inter-community connections (five motif instances) than intra-community connections (none).}
\label{Fig1_illustrative_example}
\end{figure}

For another, the quality of identified higher-order communities is easily impacted by the lack of fine-grained neighbor fuzzy community information. Specifically, community structures of real-world networks are often characterized by highly similar motif connections of bridge nodes to multiple communities and dense inter-community motif connections. Such ambiguous and complex structures can easily lead to misassigning bridge nodes and overly fragmenting communities, as shown in Fig.~\ref{Fig1_illustrative_example}. Since the community assignment of a node in previous HCD methods largely depends on the number of motif connection with communities, a node (e.g., node 4) engaging equally in motifs across multiple communities may lack adequate reference and be randomly assigned. Additionally, insufficient global optimization induced by limited network topology information often results in unreasonably fragmented communities (e.g., $C_{3}$), which usually have denser inter-community connections than intra-community ones. Merging these communities can generally improve partition quality and approximate the global optimum.

To address these issues, first, we introduce a novel concept of higher-order fuzzy memberships, which quantify the partial belongingness (i.e., membership grades) of all motif instances to crisp higher-order communities. In particular, these membership grades are calculated based on generated crisp higher-order partitions, and extend the binary membership value $\left \{0,1 \right \}$ to the continuous range of $\left [0,1 \right ]$. In this way, the degree of partial topological or functional belonging of a motif instance to multiple communities can be more accurately quantified. These higher-order fuzzy memberships not only facilitate an understanding of fuzzy and overlapping characteristics of higher-order community structures, but also provide richer and more fine-grained higher-order fuzzy community information for deep network analysis applications.

Second, the calculated higher-order fuzzy memberships are employed to enhance motif-based optimization via a general framework, called fuzzy memberships assisted motif-based evolutionary modularity (FMMEM). The main idea is to integrate the higher-order fuzzy memberships as reference assistance for solving misassigning bridge nodes and overly fragmenting communities, thereby improving the quality and accuracy of identified crisp higher-order communities, especially in ambiguity structures. In particular, considering that modularity optimization is generally NP-hard in mathematics, evolutionary algorithms (EAs) are employed as baselines for enhancing global optimization. The main contributions of this article are summarized as follows.

1) Higher-order fuzzy memberships are defined for quantifying membership grades within interval $\left [0,1 \right ]$ of motifs to crisp higher-order communities, providing richer and more fine-grained higher-order fuzzy community information. These fuzzy memberships enable deep understanding of crisp higher-order communities, and facilitate precise topology and functional analysis in other network applications.

2) A general framework FMMEM is proposed, where higher-order fuzzy memberships are employed to improve quality of optimized higher-order partitions. A fuzzy memberships-based neighbor community modification (FM-NCM) strategy corrects misassigning bridge nodes to improve partition quality. Additionally, a fuzzy memberships-based local community merging (FM-LCM) strategy combines excessively fragmented communities, thereby enhancing local search capabilities to approximate the optimum.

3) Experimental results indicate the effectiveness of higher-order fuzzy memberships in accurately reflecting partial belongingness of motifs to crisp higher-order communities. Furthermore, the proposed FMMEM framework demonstrates superior quality and accuracy compared to state-of-the-art HCD methods in both synthetic and real-world datasets, particularly those with ambiguous structures.

The rest of this study is organized as follows. Section \ref{section2} briefly reviews related work on higher-order community detection. Section \ref{section3} details the introduced higher-order fuzzy memberships. Section \ref{section4} elaborates on the proposed FMMEM framework. In Section \ref{section5}, empirical results to verify the effectiveness of FMMEM are reported. Finally, the conclusion and future prospects are given in Section \ref{section6}.

\section{Related Work}
\label{section2}

The idea of higher-order community detection based on network motifs was first proposed by Arenas et al. \cite{Arenas2008Motif, Serrour2011Detecting}. In recent years, higher-order community detection has attracted significant attention. According to the methodology of utilizing network motifs, we roughly classify current approaches into five categories: spectral clustering, local expansion, label propagation, connection enhancement, and optimization.

Motif-based spectral clustering: In the initial stages, motif-based spectral clustering methods are generally adopted. They typically utilize graph embedding techniques (e.g., spectral \cite{Benson2016Higher}, force-directed \cite{Lim2016Motif}) to map nodes into a low-dimensional space, and then employ traditional clustering algorithms (e.g., k-means) to divide the embedded nodes into clusters. Typical algorithms include ME+k-means \cite{Benson2016Higher}, TECTONIC \cite{Tsourakakis2017Scalable} and LinLog-motif \cite{Lim2016Motif}. Such methods can be scaled to large-scale networks as well as directed, signed and weighted networks, but usually find near-optimal clusters.

Motif-based local expansion: To address the difficulty of obtaining global information of networks, motif-based local expansion and optimization methods are proposed. They usually use local quality scores (e.g., local motif rate \cite{ma2019local}) to measure motif density in clusters, and combines the local expansion optimization within neighborhood of motif seeds to obtain motif-based communities. Typical algorithms include MLEO \cite{ma2019local} and MAPPR \cite{Yin2017Local}. Such methods do not require the topology information of the entire network, thus they typically run faster than global community detection methods; however, they are generally sensitive to the initial motif seeds.

Motif-based label propagation: For combining motifs with traditional community detection methods, motif-based label propagation methods have been developed. They usually construct a motif-based hypergraph to encode the higher-order connections and build a re-weighted network. Based on this, novel label updating and propagation strategies are designed to discover motif communities. Typical algorithms include MWLP \cite{li2020community} and MELPA \cite{Li2021MotifbasedEL}. Such methods are usually efficient and computationally feasible for large-scale networks, but due to the inherent randomness in the label propagation process, they may suffer from the problem of label oscillation.


Motif-based connection enhancement: motif-based connection enhancement is another typical method that integrates motifs with traditional community detection algorithms. They typically construct a motif-based hypergraph and strengthen the intracommunity connectivity structure by leveraging the motif-based connection structure. Subsequently, motif-based communities are identified within the connection enhanced hypergraph to improve quality. Typical algorithms, such as EdMot \cite{li2019EdMot}, can be easily extended to various network types (e.g., adversarial multiview) for real-life applications.

Motif-based optimization: Recently, motif-based optimization methods have been explored to more effectively utilize topological structure information and enhance global convergence performance. They model HCD as a global optimization problem based on a specific motif-based objective (e.g., motif conductance \cite{Pizzuti2017An}), and employ heuristics and EAs (e.g., genetic algorithm \cite{Pizzuti2017An}) for solving. Typical algorithms include HM-Modularity \cite{Huang2021HM-Modularity} and MotifGA \cite{Pizzuti2017An}. Such methods have advantages in approximating the global optimum, and can usually find motif-based communities with high quality and accuracy.   

The aforementioned approaches have demonstrated the effectiveness in discovering higher-order communities. Despite the success, these approaches mainly focus on discovering coarse-grained crisp higher-order community structures, and none can provide fine-grained higher-order fuzzy community information. Moreover, it is discovered that ambiguous community structures of real-life networks easily lead to these approaches misassigning nodes and overly fragmenting communities, thereby compromising the quality of identified higher-order partitions.

\section{Higher-order Fuzzy Memberships}
\label{section3}

\subsection{Preliminaries of Lower-order Fuzzy Memberships}
\label{section3_1}

To our knowledge, no methods can provide detailed higher-order fuzzy memberships. Existing lower-order fuzzy memberships \cite{Yazdanparast2021Soft, su2015quadratic, Xiao2021Fuzzy} are not suitable for higher-order community structures. Specifically, lower-order fuzzy memberships quantify the partial belongingness of nodes to lower-order crisp communities formed at edge level \cite{gregory2011fuzzy}. These fuzzy memberships provide richer community structure information in a more fine-grained manner  \cite{Havens2013A, Roy2021NeSiFCNS}. Conceptually, the lower-order fuzzy membership of a node contains membership grades to each community, valued continuously between $[0, 1]$. The membership grade expands the binary membership value $\left \{0,1\right \}$ corresponding to lower-order crisp community structures, thus offering a more detailed and accurate depiction of a node realistic partial community affiliations \cite{Biswas2018FuzAg, Roy2021NeSiFCNS}. Lower-order fuzzy memberships significantly promote understanding of topology and functional characteristics of lower-order crisp communities in real-life networks \cite{Xiao2021Fuzzy}.

Formally, given a network $G$$=$$(V, E)$, where $V$ is a set of vertices ($n$$=$$|V|$) and $E$ is a set of edges ($m$$=$$|E|$). The lower-order fuzzy membership of the $i$th node is denoted by $\textbf{\emph{u}}_{i}$ = $\left [u_{1i}, u_{2i}, \cdots, u_{ki}, \cdots, u_{ci} \right ]$, where $c$ denotes the number of lower-order crisp communities. Here, $u_{ki}$ indicates the membership grade of the $i$th node to the $k$th community, with values ranging from 0 to 1. Particularly, the sum of membership grades for a single node typically equals ``1", and the highest membership grade often determines the crisp community affiliation \cite{Havens2013A, su2015quadratic}. The more neighbor nodes (or edge connections) a node has within a community, the higher the membership grade to the community. 

Currently, the concept of fuzzy memberships is seldom utilized to analyze higher-order community structures, leaving the partial affiliations of motif instances to higher-order crisp communities mostly unexplored. Existing lower-order fuzzy memberships, primarily focus on lower-order crisp communities based on edge-level connectivity patterns, cannot effectively reflect the fuzzy membership relationships between motifs and communities based on motif-level higher-order connectivity patterns. Consequently, valuable higher-order fuzzy topology information regarding higher-order crisp community structures remains largely unknown.

\subsection{Constructing Higher-order Fuzzy Memberships}
\label{section3_2}
To enhance the richness and detail of higher-order fuzzy topology information for higher-order crisp community structures, we introduce a novel concept: higher-order fuzzy memberships. Unlike lower-order fuzzy memberships, higher-order fuzzy memberships quantify the partial affiliations of any motif to all higher-order crisp communities. For example, in a scientist collaboration network, higher-order fuzzy memberships can measure how a research team (formed as a motif) is partially involved in various academic organizations. Therefore, higher-order fuzzy memberships accurately depict the realistic topology connectivity and functional participation of motifs to higher-order crisp communities in real-life networks.

Formally, given a specific motif $M$, we define higher-order fuzzy memberships and related concepts as follows.
 
\textit{Definition 1 \bf{(Higher-order Fuzzy Memberships of Motifs)}}: The higher-order fuzzy membership for an instance $H$ of motif $M$ is defined as $\textbf{\emph{u}}_{H}$ = $\left [ u_{1H}, u_{2H}, \cdots, u_{kH}, \cdots, u_{cH} \right ]$, where $c$ denotes the number of higher-order crisp communities, and $u_{kH}$ indicates the higher-order membership grade of $H$ to the $k$th community. Particularly, $u_{kH}$ is positive if the $k$th community is among the neighbor communities $N_{M}^{C}(H)$ of $H$, and zero otherwise. Here, $N_{M}^{C}(H)$ defines the set of motif-based neighbor communities of nodes in $H$, that is
\begin{equation}
N_{M}^{C}(H)=\left \{ C|C\in N_{M}^{C}(v),v \in H \right \}.
\label{equation1}
\end{equation}
where $N_{M}^{C} (v)$ represents the set of motif-based neighbor community of a node $v$.

\textit{Definition 2 \bf{(Motif-based Neighbor Community)}}: Given a node $v\in V$ and a specific motif $M$, if $v$ belongs to community $C_{v}$ and there exists another community $C_{u}$ that contains any motif-based neighbor node of $v$, we define community $C_{u}$ as a motif-based neighbor community of $v$ based on $M$. The set of motif-based neighbor community of $v$ can be expressed as $N_{M}^{C} \left ( v\right ) =\left \{ C_{u}| \left ( u \in C_{u}\right ) \wedge \left ( u \in N_{M} \left ( v\right )\right ), C_{v} \neq C_{u} \right \}$, where $N_{M} \left ( v\right )$ defines the set of motif-based neighbors of a node $v$.

\textit{Definition 3 \bf{(Motif-based Neighbor Node)}}: Given a node $v\in V$ and a specific motif $M$, if there exists another node $u \in V$ ($u \neq v$) that $u$ and $v$ co-occur in any motif instance of $M$, we define node $u$ as a motif-based neighbor node of $v$ based on $M$, regardless of whether there is an edge between them. The set of motif-based neighbors of node $v$ can be expressed as $N_{M} \left ( v\right ) =\left \{u | v\in H ,u\in H , H \cong M , v \neq u \right \}$.

In particular, the evaluation of higher-order membership grades of a motif instance $H$ depends on the closeness of motif connections. Specifically, a greater number of motif instances connected to nodes within $H$ in the $k$th community increases the value of $u_{kH}$. Given a motif instance $H$, its membership grade to the $k$th community that belongs to $N_{M}^{C}(H)$, is calculated as the sum of the higher-order membership grades for each node $v$ in $H$, that is
\begin{equation}
u_{kH}=\frac{\textstyle\sum_{v\in H}u_{kv}}{\left \|{V}_{H} \right \|},
\label{equation2}
\end{equation}
where $u_{kv}$ denotes the higher-order membership grade of a node $v$ in $H$ to the $k$th community, and $\left \|{V}_{H} \right \|$ denotes the total number of nodes in $H$. The value of $u_{kH}$ ranges from 0 to 1, higher values signify stronger membership. The total membership grades of $H$ to all communities sum to ``1".

\textit{Definition 4 \bf{(Higher-order Fuzzy Memberships of Nodes)}}: The higher-order fuzzy membership of a node $v$ is defined as  $\textbf{\emph{u}}_{v}$ = $\left [ u_{1v},u_{2v},\cdots,u_{kv}\cdots,u_{cv} \right ]$, which quantifies the higher-order membership grades of node $v$ to all neighbor communities $N^{C}(v)$. Here, $N^{C}(v)$ denotes the set of communities to which the neighbors of $v$ belong, that is
\begin{equation}
N^{C}(v)=\left \{ C|u \in C \land u \in N_{v} \right \},
\label{equation3}
\end{equation}
where $N_{v}$ is the edge-based neighbor set of $v$. 

The evaluation of higher-order fuzzy memberships of nodes depends on the connection strengths at both motif (higher-order) and edge (lower-order) levels, thus more comprehensively and accurately revealing the realistic topology or functional memberships of a node within higher-order crisp communities. Specifically, for a node $v$, its membership grade to a neighbor community ${C}_{k}$ is defined as the ratio of the attraction of ${C}_{k}$ to $v$ relative to the total attraction from all neighbor communities in ${N^{C}(v)}$ to $v$, that is
\begin{equation}
u_{kv}=\frac{Attr(v,{C}_{k})}{\textstyle\sum_{C \in N^{C}(v)}Attr(v,{C})},
\label{equation4}
\end{equation}
where $Attr(v,{C})$ denotes the attraction from any neighbor community ${C\in N^{C}(v)}$ to $v$, calculated as the difference between the total connection strength from all nodes in $C$ to $v$ and its mathematical expectation, that is
\begin{equation}
Attr(v,{C})=max\left \{\textstyle\sum_{u\in C}(w_{uv}-\frac{{S}_{u}*{S}_{v}}{2W}),0 \right \},
\label{equation5}
\end{equation}
where $\textstyle\sum_{u\in C}w_{uv}$ and $\textstyle\sum_{u\in C}\frac{{S}_{u}*{S}_{v}}{2W}$ denote the actual connection strength and its mathematical expectation between node $v$ and all nodes in $C$, repectively. Additionally, $w_{uv}$ represents the multi-order weight of the edge ($u$,$v$), $W$ represents the total of all multi-order edge weights in $G_{W}$, and ${S}_{u}$ and ${S}_{v}$ denote the strengths of nodes $u$ and $v$, respectively. A higher $Attr(v,{C})$ value suggests greater topological attractiveness of the neighbor community ${C}$ to $v$. Furthermore, the membership grade $u_{kv}$ ranges from 0 to 1, and the total membership grades of $v$ to all neighbor communities sum to ``1".

\subsection{Effectiveness of Higher-order Fuzzy Memberships}
\label{section3_3}

\begin{figure}[ht]
\centering
\includegraphics[width=0.40\textwidth]{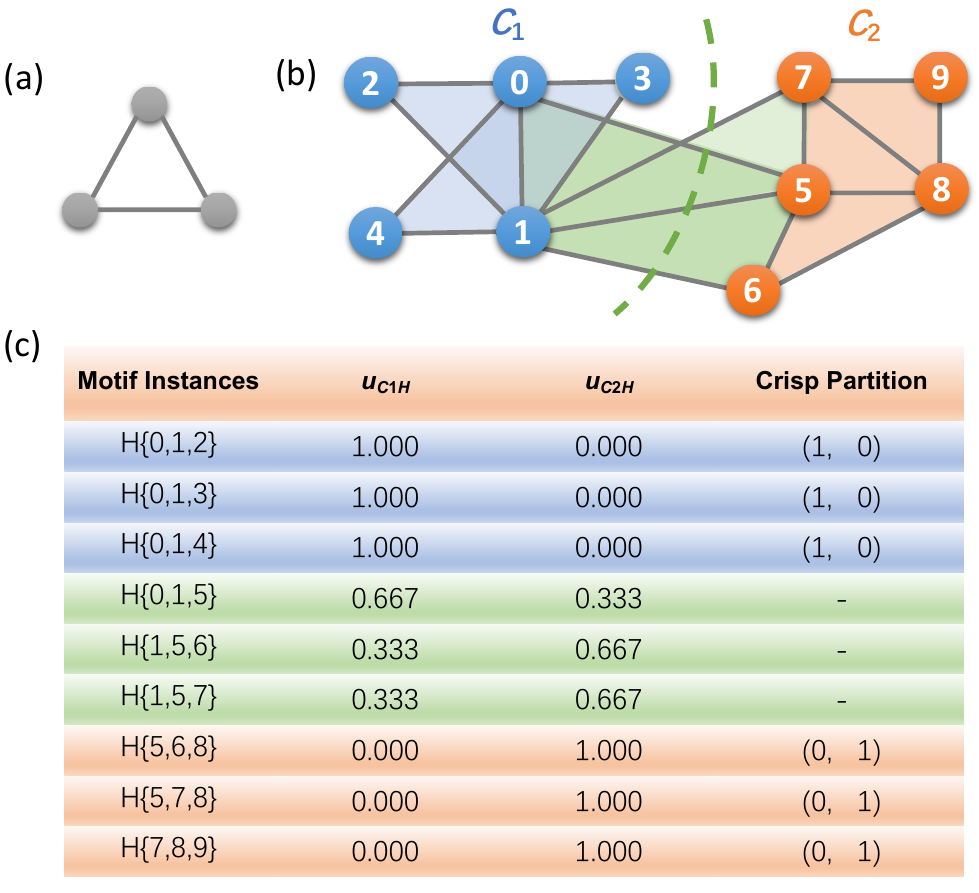}
\caption{The example of higher-order fuzzy memberships for the ground-truth communities of a simple network. (a) A triangular motif widely existed in the network; (b) ground-truth higher-order crisp partition of the network containing two higher-order communities: ${C}_{1}$ (blue nodes) and ${C}_{2}$ (orange nodes); and (c) higher-order fuzzy memberships of all nine motif instances, i.e., $H\left \{0,1,2 \right\}$, $H\left \{0,1,3 \right\}$,$\cdots$,$H\left \{7,8,9 \right\}$, accurately reveal their fine-grained partial memberships for the ground-truth higher-order crisp communities.}
\label{Fig2_motifs_fuzzy_memberships}
\end{figure}

First, the higher-order fuzzy memberships effectively enrich fine-grained higher-order fuzzy community information. The verification is performed on a typical network with a known higher-order crisp partition \cite{Benson2016Higher} in Fig.~\ref{Fig2_motifs_fuzzy_memberships}. Utilizing a triangular motif as depicted in Fig.~\ref{Fig2_motifs_fuzzy_memberships}(a), the nodes are categorized into two higher-order crisp communities: ${C}_{1}$ (blue nodes) and ${C}_{2}$ (orange nodes), shown in Fig.~\ref{Fig2_motifs_fuzzy_memberships}(b). As defined, the higher-order fuzzy memberships of all nine motif instances to two higher-order crisp communities are quantified in Fig.~\ref{Fig2_motifs_fuzzy_memberships}(c).

The quantified higher-order fuzzy memberships in Fig.~\ref{Fig2_motifs_fuzzy_memberships}(c) can accurately reveal the fine-grained partial memberships for all nine motif instances to the ground-truth higher-order crisp communities in Fig.~\ref{Fig2_motifs_fuzzy_memberships}(b). Specifically, the first group motif instances, i.e.,$H\left \{0,1,2 \right\}$, $H\left \{0,1,3 \right\}$ and $H\left \{0,1,4 \right\}$, share identical fuzzy memberships with membership grade of ``1" to ${C}_{1}$. The second group motif instances, i.e.,$H\left \{5,6,8 \right\}$, $H\left \{5,7,8 \right\}$ and $H\left \{7,8,9 \right\}$, have membership grade of ``1" to ${C}_{2}$. Particularly, the third group motif instances, i.e., $H\left \{0,1,5 \right\}$, $H\left \{1,5,6 \right\}$ and $H\left \{1,5,7 \right\}$, exhibit ambiguous membership grades to ${C}_{1}$ and ${C}_{2}$ within the interval $\left [0,1 \right ]$, indicating their partial community affiliations with varying degrees. The results prove that the constructed higher-order fuzzy memberships can offer accurate higher-order fuzzy topology information for the ground-truth community structures.

Second, fine-grained neighbor fuzzy community information can serve as a detailed reference to enhance the quality of identified higher-order communities. Specifically, higher-order fuzzy memberships can be used to relocate misassigned bridge nodes within neighbor communities with stronger multi-order topological connections, as illustrated in Fig.~\ref{Fig3_FM_NCM}, thereby improving the quality of optimized higher-order partitions. Additionally, they can also be adopted to precisely identify excessively fragmented communities and their neighbor communities with high membership relationship, as illustrated in Fig.~\ref{Fig4_FM_LCM}, thereby enhancing local community merging and approximating the global optimum.

\section{Fuzzy Menberships Assisted Motif-based Evolutionary Modularity Optimization}
\label{section4}

\subsection{Definition of Higher-order Community Detection}
\label{section4_1}

Given a network $G$$=$$(V, E)$ with node set $V$($\left | V\right |$$=$$n$) and edge set $E$($\left | E\right |$$=$$m$), network motifs in $G$ are defined as interconnected subgraphs occurring at numbers that are significantly higher than those in randomized networks preserving the same degree distribution \cite{Milo2002Network}. Formally, a network motif can be expressed as $M\left (p,q\right )$$=$$\left \{V_{M},E_{M}\right \}$, where $V_{M}$$\subseteq$$V$ and $E_{M}$$\subseteq$$E$ represent the node set with $p$ nodes and edge set with $q$ edges involved in the motif $M$, respectively \cite{li2019EdMot}. For example, a triangular motif with 3 nodes and 3 edges can be represented as $M\left (3,3\right )$. For convenience, a network motif is usually expressed simply as $M$, and a subgraph in $G$ that is isomorphic to the motif $M$ is denoted as a motif instance, that is $H$$\cong$$M$ \cite{Huang2021HM-Modularity}. The set of motif instances of $M$ in graph $G$ can be defined as
\begin{equation}
S\left ( M\right )=\left \{H | H \subset G\ \wedge H \cong M \right \}.
\label{equation6}
\end{equation}

Network motifs are rich in forms of structure expression, representing different interactions among multiple nodes and generally corresponding to different functional characteristics \cite{Benson2016Higher}. Particularly, the 3-node triangular motif $M\left (3,3\right )$ is both prevalent and crucial in real-life networks, such as social and biological networks. Therefore, this motif is extensively employed in current HCD methods \cite{li2019EdMot, ma2019local, Huang2020Efficient} and is also the primary focus of our work.

Higher-order community detection with a specific motif can be viewed as a process of finding a higher-order partition, where motif connections within each community should be as dense as possible and between communities as sparse as possible \cite{Huang2020Efficient}. For a specific motif $M$ in network $G$, a motif adjacency matrix $\textbf{\emph{W}}^{M}$$=$$\left [w_{ij}^{M} \right ]^{n \times n}$ is usually constructed first. Each element $w_{ij}^{M}$$=$$\sum_{H}^{} \left \{ H | v_{i}\in  H,v_{j}\in H, H \cong M \right \}$  represents the number of motif instances containing nodes $v_{i}$ and $v_{j}$. The greater the value of $w_{ij}^{M}$, the closer the higher-order connections between the nodes. Based on motif adjacency matrix $\textbf{\emph{W}}^{M}$, a higher-order partition can be obtained by dividing network $G$ into a family of node groups. The number of motif instances in the same group is large, while that among different groups is small \cite{Pizzuti2017An}. Formally, a higher-order partition can be expressed as $\textbf{\emph{C}}$$=$$\{C_{1},C_{2},\cdots, C_{c}\}$, which meets the following basic requirements:
\begin{equation}
\left\{\begin{matrix}
V=\bigcup_{i=1}^{c}C{_{i}}, \\
C_{i} \subset V, \: i=\left \{1,2,\cdots,c\right \},\\
C_{i} \neq \varnothing,\:  i=\left \{1,2,\cdots,c\right \},\\
C_{i} \neq C_{j}, \forall i\neq j, \:  i,j \in \left \{1, 2, \cdots, c\right \}, \\
\end{matrix}\right.
\label{equation7}
\end{equation}
where $C_{i}$ represents a group of nodes involved in the $i$th community, and $c$ represents the number of communities. In the higher-order partition $\textbf{\emph{C}}$, a motif instance $H$ only belongs to one community, that is, $C_{i}\cap C_{j}$$=$$\varnothing$, $\forall i$$\neq$$j$, $i,j $$\in$$ \left \{1,2,\cdots,c\right \}$.

\subsection{Motif-based Modularity Optimization Model}
\label{section4_2}

A motif-based modularity optimization model is developed to effectively utilize motif structural information, while preserving the lower-order topology, such as nodes and edges. The core idea is to integrate the number of motif instances as weights between nodes, transforming the HCD problem into traditional CD on a weighted network, and constructing it into a weighted modularity optimization problem for solving.

Second, a weighted graph $G_{W}$ is constructed by incorporating multi-order topology information as weights between nodes, considering both motif (higher-order) and edge (lower-order) connections. Given a network $G$$=$$(V, E)$ with a specific motif $M$, each non-zero element $w_{ij}^{M}$ in the motif adjacency matrix $\textbf{\emph{W}}^{M}$$=$$\left [w_{ij}^{M} \right ]^{n \times n}$ serves as the higher-order weight of the edge $(v_{i},v_{j})$, indicating the higher-order compactness between motif-based neighbors $v_{i}$ and $v_{j}$. Furthermore, by summing the higher-order weights with the non-zero elements in the lower-order adjacency matrix $\textbf{\emph{A}}$$=$$\left [a_{ij}\right ]^{n \times n}$, multi-order topological weights are formed, which are used to construct the weighted network $G_{W}$$=$$(V, E, \textbf{\emph{W}})$. In particular, $G_{W}$ retains the complete lower-order topology (nodes and edges) of $G$, thus greatly reducing the occurrence of isolated nodes and fragmented components.

Third, in order to obtain higher-order communities, a modularity optimization model is established on the weighted network $G_{W}$. Ideally, a higher-order community partition should maximize intra-community motif connections and minimize inter-community ones \cite{Benson2016Higher, Tsourakakis2017Scalable, Huang2020Efficient}. In other words, the motif-based weights within each community should be maximized, while minimized between communities, achievable through weighted modularity optimization. Furthermore, the multi-order weights, which reflect the closeness of topological connections between nodes, aid in overcoming resolution limitations in modularity optimization \cite{Newman2004Analysis, Cao2016Weighted}, thereby enhancing detection performance.

Based on the above analysis, the motif-based modularity optimization model is constructed as a global maximization problem and formally defined as
\begin{equation}
F({\textbf{\emph{C}}}^*) = max (F(\textbf{\emph{C}})) ,
\label{equation8}
\end{equation}
where $\textbf{\emph{C}}$ stands for a candidate motif-based community partition in the search space $\bf{\Omega}$. $F$$:$$\bf{\Omega}$$\to$$\textbf{\emph{R}}$ represents an objective function used to evaluate the quality of motif-based communities. Here, weighted modularity $Q_{W}$ \cite{Newman2004Analysis} is employed as the objective function, which is defined as
\begin{equation}
Q_{W} = \frac{1}{2W}\sum_{ij}\left (w_{ij}-\frac{s_{i}s_{j}}{2W} \right )\delta \left (i,j \right ),
\label{equation9}
\end{equation}
where $W$$=$$\frac{1}{2}\sum_{ij}^{}w_{ij}$ is the sum of weights of all the edges in $G_{W}$. The strengths of $v_{i}$ and $v_{j}$ (i.e., $s_{i}$ and $s_{j}$), equal to the sum of weights $\sum_{j\in N_{i}}^{}w_{ij}$ and $\sum_{i\in N_{j}}^{}w_{ij}$, respectively. $N_{i}$ and $N_{j}$ denote the edge-based neighbor sets of $v_{i}$ and $v_{j}$, respectively. $\delta \left (i,j\right )$ yields one if nodes $v_{i}$ and $v_{j}$ belong to the same community, zero otherwise. The community partition $\textbf{\emph{C}}^{*}\in\bf{\Omega}$ with the highest $Q_{W}$ value is regarded as the optimal partition of the weighted network $G_{W}$, which is also the optimal higher-order community partition based on motif $M$ for the original network $G$.

\subsection{General Framework of FMMEM}
\label{section4_3}

Modularity-based optimization is generally proven to be a NP-hard problem in mathematics and therefore extremely difficult to solve \cite{Pizzuti2018Evolutionary}. Evolutionary algorithms (EAs) possess strong global optimization capabilities for solving complex problems and approximating the global optimum, making them widely used in modularity-based optimization \cite{Zhang2019Application}. Compared to traditional deterministic optimization algorithms and heuristics, EAs present a number of advantages \cite{Pizzuti2018Evolutionary}. First, they are global search algorithms with the characteristic of population-based naturally parallel and the basic principle of self-organization. Second, network structure information can be utilized in the algorithms to enhance the exploration and exploitation of search space. Third, the number of communities can be automatically determined during the optimization.

Although the EAs-based modularity optimization is effective in community detection, it has rarely been employed to discover higher-order communities. The primary challenges include three aspects: (i) motif information should be effectively used to construct a higher-order modularity function as well as an optimization model; (ii) EAs should be properly incorporated in the motif-based modularity optimization model; and (iii) network topology structural information should be effectively utilized in key evolutionary operators, providing search directions and helping to approximate the optimum.

Based on the analysis, a general framework of fuzzy memberships assisted motif-based evolutionary modularity (FMMEM), is proposed to approximate the optimal higher-order community partition. In particular, FMMEM adopts a EAs-based motif modularity optimization framework, which is presented in Algorithm \ref{FFMEM_algorithm}. The core idea is to utilize the higher-order fuzzy memberships as reference to assist the motif-based evolutionary modularity optimization, via addressing the two typical problems of misassigning nodes and overly fragmenting communities, thereby enhancing the quality of identified higher-order communities, particularly in ambiguous structures.

\renewcommand{\algorithmicrequire}{\textbf{Input:}}
\renewcommand{\algorithmicensure}{\textbf{Output:}}
\begin{algorithm}[!h]
\caption{General Framework of FFMEM.}
\begin{algorithmic}[1]
\REQUIRE $G$: original network; $\textbf{\emph{A}}$: adjacency matrix; $M$: motif of interested; $t_{max}$: maximum number of iterations; $\text{\emph{NP}}$: population size;  \\
\ENSURE $\textbf{\emph{C}}^{*}$: community partition;
   \STATE $\textbf{\emph{W}}^{M}$, $G_{W}$ $\leftarrow$ generate motif adjacency matrix of $G$ with motif $M$, and construct the weighted network with multi-order topological weights $\textbf{\emph{W}}$; // Section V-A;
   \STATE $\textbf{\emph{P}}$ $\leftarrow$ initialize the population $\textbf{\emph{P}}$=$\left [ \textbf{\emph{X}}_{1}, \textbf{\emph{X}}_{2},\cdots, \textbf{\emph{X}}_\text{\emph{NP}} \right ]$;
   \STATE \textbf{for} $iter$ = 1 to $t_{max}$ \textbf{do}
   \STATE $\quad$ $\textbf{\emph{P}}_{OPT}$ $\leftarrow$ update $\textbf{\emph{P}}$ by a specific baseline EA;  
   \STATE $\quad$ $\textbf{\emph{P}}_{NCM}$ $\leftarrow$ update $\textbf{\emph{P}}_{OPT}$ by the FM-NCM strategy; // Section V-D; 
   \STATE $\quad$ $\textbf{\emph{P}}_{LCM}$ $\leftarrow$ update $\textbf{\emph{P}}_{NCM}$ by the FM-LCM strategy; // Section V-E; \\
   \STATE $\quad$ $\textbf{\emph{P}}$ $\leftarrow$ replace $\textbf{\emph{P}}$ with $\textbf{\emph{P}}_{LCM}$;
   \STATE \textbf{end for}
   \STATE $\textbf{\emph{C}}^{*}$ $\leftarrow$ the individual in $\textbf{\emph{P}}$ with the largest $Q_{W}$ value;\\
   \STATE \textbf{return} $\textbf{\emph{C}}^{*}$
\end{algorithmic}
\label{FFMEM_algorithm}
\end{algorithm}

The general framework of FMMEM is constructed in three steps: weighted network construction, population initialization, and evolutionary modularity maximization.
\begin{itemize}
   \item [1)] 
   Weighted Network Construction: Construct the motif adjacency matrix $\textbf{\emph{W}}^{M}$, reflecting the number and distribution of a specific motif $M$. Subsequently, develop a weighted network $G_{W}$ incorporating both motif and edge-based multi-order topology information as weights. 
   \item [2)]
   Population Initialization: Establish an initial population $\textbf{\emph{P}}$ consisting of $N_{pop}$ individuals, using a label-based encoding schema \cite{Pizzuti2018Evolutionary, Zhang2019Application}. This schema allows the algorithm to automatically determine the number of communities without requiring a decoding process. 
   \item [3)]
   Evolutionary Modularity Maximization: Iteratively evolve population $\textbf{\emph{P}}$ using a specific baseline EA and two strategies (i.e., FM-NCM and FM-LCM) to maximize $Q_{W}$. This is the key step of FMMEM to approximate the optimal higher-order community partition.
\end{itemize}

Specifically, each iteration of FMMEM involves three key operations: (i) evolving the population to generate high-quality offspring individuals using evolutionary operators from the baseline EA, (ii) constructing higher-order fuzzy memberships to modify misassigned bridge nodes within neighbor communities, enhancing higher-order partition quality via the FM-NCM strategy, and (iii) utilizing higher-order fuzzy memberships to merge excessively fragmented communities, improving local search ability and approximate the global optimum through the FM-LCM strategy.

\subsection{Fuzzy Memberships based Neighbor Community Modification}
\label{section4_4}

Node community mismatches frequently arise during motif-based modularity optimization, particularly in EA-based random searches. For instance, nodes with motif connections might be allocated to a community lacking motif instances, or they could be less densely connected to the community’s motifs compared to those in other communities. Moreover, in cases where community structures are complex and ambiguous, the topological relationships of bridge nodes with several neighbor communities may become indistinct, thereby heightening the risk of community mismatches.

To address the issue, we propose a fuzzy memberships-based neighbor community modification (FM-NCM) strategy. The core concept involves using constructed higher-order fuzzy memberships to provide fine-grained neighbor fuzzy community information for rectifing mismatched bridge nodes, thereby improving the quality of higher-order partitions. Unlike existing strategies that primarily rely on coarse-grained discrete topology features (e.g., number of motif instances and edge connections), FM-NCM incorporates fuzzy memberships of motifs to neighbor communities as more detailed assistant information. These fine-grained fuzzy memberships increase the probability that a mismatched bridge node will be relocated to a local optimal neighbor community with the strongest multi-order topological connections.

\begin{figure}[ht]
\centering
\includegraphics[width=0.45\textwidth]{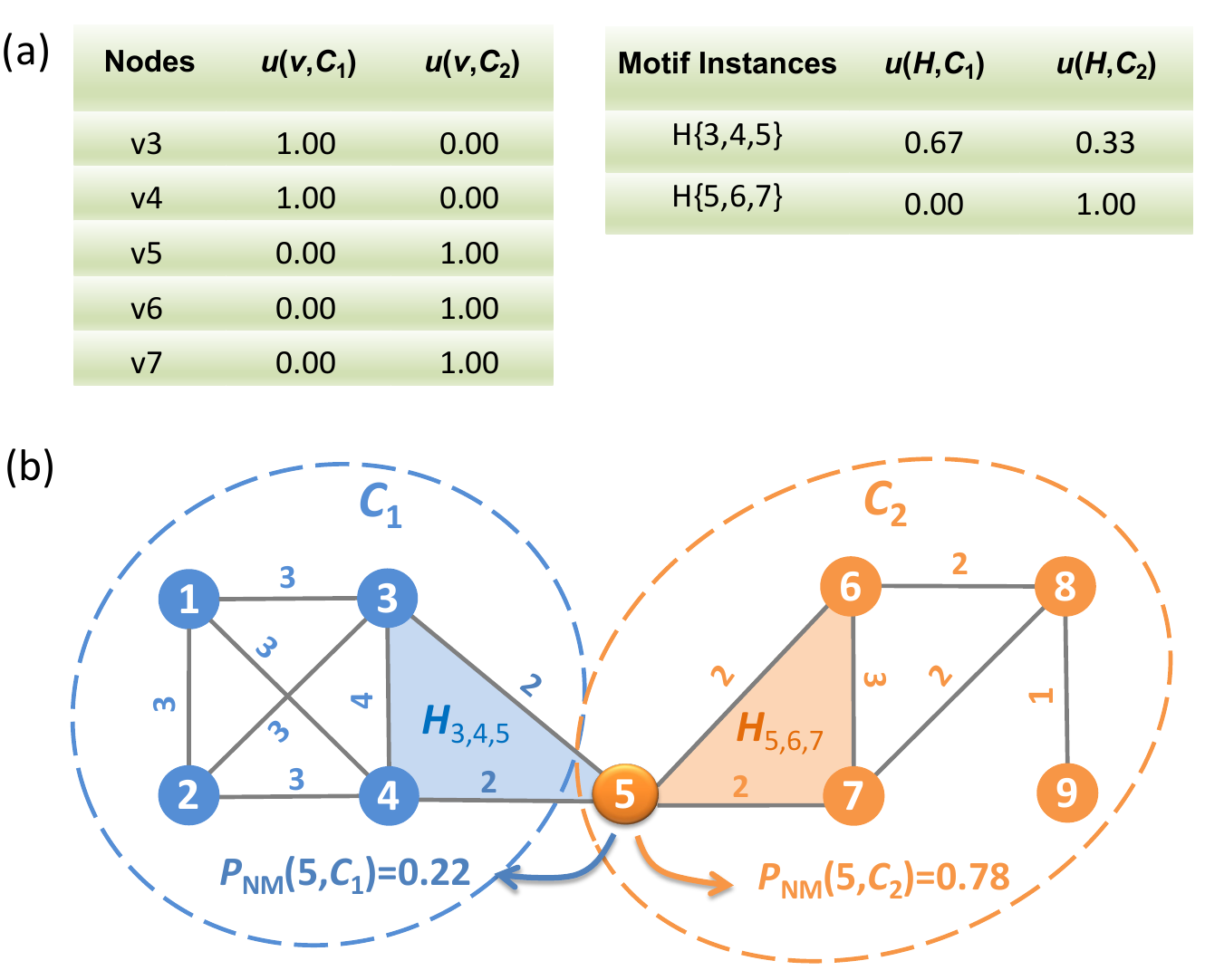}
\caption{An illustration of the FM-NCM strategy applied to a basic network comprising two communities (i.e., $C_{1}$ and $C_{2}$) using a 3-node triangular motif. (a) Fuzzy memberships for target node 5 and four neighbor nodes, and fuzzy memberships for two motif instances involving target node 5; (b) Normalized migration probabilities for node 5 to neighbor communities $C_{1}$ and $C_{2}$, and the resulting community migration.} 
\label{Fig3_FM_NCM}
\end{figure}

The FM-NCM strategy primarily consists of three steps: selecting target nodes, calculating migration probabilities, and migrating mismatched nodes, which are elaborated using an illustrative example depicted in Fig.~\ref{Fig3_FM_NCM}. In the network, which features a higher-order partition with two communities, $C_{1}$ and $C_{2}$, by using a 3-node triangular motif, we initially construct a target node set, $V_{obj}$, comprising three bridge nodes $\left \{3,4,5 \right \}$. These bridge nodes, connecting to the communities beyond their assigned ones, are prone to misclassification in community assignments. 

Second, for each target node $v \in V_{obj}$ (e.g., node 5) and its neighbor community $C_{k}$ (e.g., $C_{1}$ and $C_{2}$), we compute the membership grade $u(v,{C}_{k})$ of node $v$ to $C_{k}$, and the membership grade $u(H,{C}_{k})$ of each participated motif instance $H$ to ${C}_{k}$, as shown in Fig.~\ref{Fig3_FM_NCM}(a). The aim is to accurately assess the topological connection strength of target node $v$ with community $C_{k}$ at both lower- and higher-order levels. On this basis, a normalized migration probability of node $v$ to $C_{k}$ is defined as
\begin{equation}
{P}_{NM}(v,{C}_{k})=\frac{u(v,{C}_{k})+\textstyle\sum_{H\in{S}_{v}(M)}u(H,{C}_{k})}{1+\left \|{S}_{v}(M) \right \|},
\label{equation10}
\end{equation}
where ${S}_{v}(M)=\left \{H|v\in H,H\cong M\right \}$ represents the set of all motif instances participated by node $v$. The higher the value of ${P}_{NM}(v,{C}_{k})$, the stronger the multi-order topological connections, and the greater the probability that node $v$ will migrated to community $C_{k}$.

At last, if the community of a target node $v$ does not correspond to the one with the highest migration probability, node $v$ will randomly migrate to a neighbor community based on the respective migration probabilities. This enhances the probability that node $v$ will adjust to a neighbor community, which is more closely connected according to the multi-order topology. For instance, as shown in Fig.~\ref{Fig3_FM_NCM}(b), node 5 is migrated to $C_{2}$ with a higher migration probability. Therefore, the local search capabilities of bridge nodes and the overall quality of higher-order community partitions can be improved. 

\subsection{Fuzzy Memberships based Local Community Merging}
\label{section4_5}

Higher-order communities often show excessive fragmentation in generated partitions, especially in relatively larger networks with complex motif connections. This frequently results in numerous unreasonable smaller communities with denser inter-community motif connections. Reasons mainly include inadequate global optimization and local community search induced by limited network topology information, expecially fine-grained neighbor fuzzy community information. Merging these communities can generally improve partition quality and approximate the global optimum.

To address the issue, we propose a fuzzy memberships-based local community merging (FM-LCM) strategy. This strategy aims to utilize the constructed higher-order fuzzy memberships to offer fine-grained neighbor fuzzy community information for merging excessively fragmented communities, thereby improving the local community search towards approximating the global optimum. Unlike existing strategies \cite{Pizzuti2020MOL}, which focus solely on the number of lower-order edge connections between bridge nodes, FM-LCM employs higher-order fuzzy memberships as more detailed assistant information to identify target smaller communities requiring local merging, and incorporates edges and motifs-based connections to comprehensively assess the closeness of all nodes within neighbor communities. This provides more detailed and accurate metric information, facilitating the precise selection of the target and neighbor communities for local merging.

The FM-LCM strategy comprises three primary steps: selecting target partitions and communities, calculating multi-order topology closeness, and merging local communities. First, candidate partitions are chosen from the population, consisting of individuals with the modularity value remains unchanged for a number of iterations exceeding the threshold $\delta_{Iter}$. Particularly, if an individual has served as a target partition for $\delta_{max}$ consecutive iterations without any community merging or modularity improvement, it is deemed to have converged and is no longer considered a target partition. In each target partition, candidate communities considered for local merging are those where the intra-community nodes exhibit a certain of membership grades to any neighbor community, such as the average node membership grade exceeds 0.2.

Second, we calculate the multi-order topology closeness between community $C$ and its neighbor communities as
\begin{equation}
{T_{c}}(C,C_{k})=\textstyle\sum_{i\in C,j \in C_{k}}{T_{v}}(i,j),
\label{equation11}
\end{equation}
where $C_{k}$ denotes a neighbor community connected to $C$, ${T_{v}}$ defines the multi-order topology closeness between any two nodes within $C$ and $C_{k}$ respectively, that is
\begin{equation}
{T_{v}}(i,j)=w_{ij}-\frac{{S}_{i}*{S}_{j}}{2W},
\label{equation12}
\end{equation}
where $w_{ij}$ denotes the multi-order weight of edge (i,j), $W$ represents the total multi-order weights of all edges in $G_{W}$, and ${S}_{i}$ and ${S}_{j}$ indicate the strengths of nodes $i$ and $j$, respectively. Finally, the neighbor community $C_{k}$, which has the highest positive ${T_{c}}$ value, is chosen for local merging with $C$ to enhance the modularity quality of the target partition. The local merging process continues within the target partition’s communities until no further enhancement in the modularity $Q_{W}$ value is possible.

\begin{figure}[ht]
\centering
\includegraphics[width=0.45\textwidth]{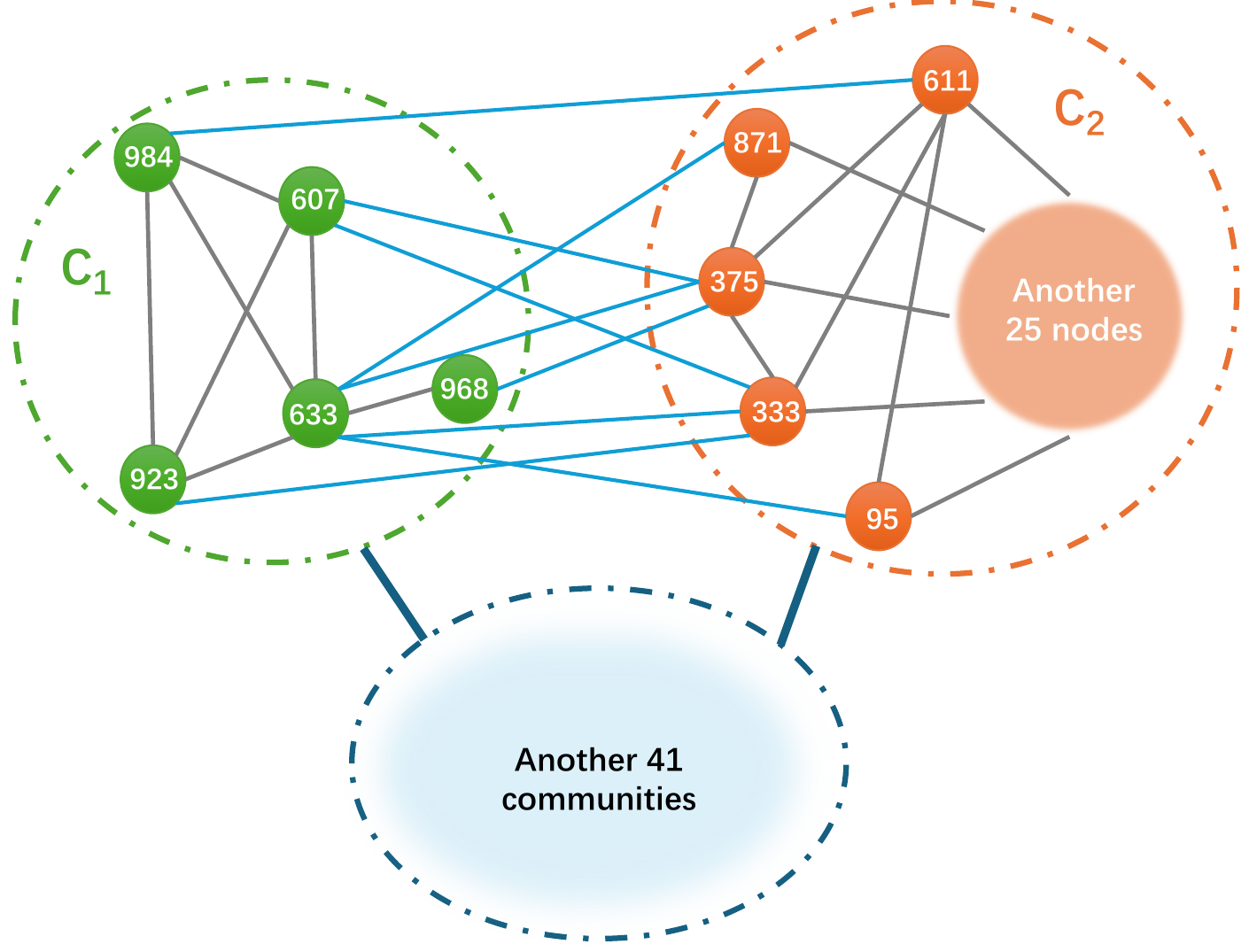}
\caption{An illustration of the FM-LCM strategy on a higher-order partition of the Email network, which include 43 communities identified using a triangular motif. A smaller community ${C_{1}}$ is chosen as a target community, and a larger neighbor community ${C_{2}}$ that has the highest multi-order topology closeness ${T_{c}}$ is selected for merging. Continuous local merging improves the modularity quality of the higher-order partition, increasing it from 0.681 to 0.699.}
\label{Fig4_FM_LCM}
\end{figure}

Figure.~\ref{Fig4_FM_LCM} shows an illustration of the FM-LCM strategy applied to a higher-order partition of the Email network, which includes 43 communities identified using a triangular motif. A smaller community ${C_{1}}$ is chosen as a target community due to the relatively high average membership grades of intra-community nodes to the neighbor communities. For instance, for nodes within ${C_{1}}$, their average membership grade to ${C_{2}}$ achieves 0.978. Additionally, ${C_{2}}$ with the highest multi-order topology closeness ${T_{c}}$ is selected for merging with ${C_{1}}$. Continuous local merging improves the modularity quality of the higher-order partition, increasing it from 0.681 to 0.699.

\subsection{Time complexity Analysis}
\label{section4_6}
The computation of FMMEM for a single network involves three main components: (i) identifying motifs \cite{Wernicke2006FANMODAT} and constructing the motif adjacency matrix $\textbf{\emph{W}}^{M}$ for a specific motif $M$, (ii) iteratively performing the evolutionary operators of the baseline EA algorithm, FM-NCM, and FM-LCM, to maximize modularity, and (iii) assessing individuals using the weighted modularity $Q_{W}$.

First, $\textbf{\emph{W}}^{M}$ can be computed in $O(n^{k})$ time \cite{Benson2016Higher}. Second, the computational complexity of evolutionary operators in typical EAs (e.g., GA) is generally $O(n)$. For the FM-NCM and FM-LCM operations, the worst-case time complexity are $O(n\left \langle d\right \rangle^{2})$ and $O(n\left \langle d\right \rangle)$, respectively, with $\left \langle d\right \rangle$ denoting the average node degree. Third, the evaluation of a motif-based partition using $Q_{W}$ incurs a worst-case computational cost of $O(n^{2})$. In summary, the time complexity of FMMEM with the 3-node triangle motif can be expressed as $O(m^{1.5})+t_{max}*N_{pop}*\left \{O(n)+O(n\left \langle d\right \rangle^{2})+O(n\left \langle d\right \rangle)+ O(n^{2}) \right \}$, and can be simplified to $O(m^{1.5})+t_{max}*N_{pop}*O(n^{2})$. Here, $t_{max}$ denotes the maximum number of iterations, while $N_{pop}$ refers to the population size. The time complexity of FMMEM is the same as other motif-based evolutionary algorithms \cite{Pizzuti2017An}.

\section{EMPIRICAL RESULTS}
\label{section5}

\subsection{Experimental Settings}
\label{section5_1}

\subsubsection{Comparison Algorithms}
The performance of FMMEM is compared with state-of-the-art higher-order community detection algorithms, namely, ME+k-means \cite{Benson2016Higher}, MLEO \cite{ma2019local}, MELPA \cite{Li2021MotifbasedEL}, EdMot \cite{li2019EdMot} and MotifGA \cite{Pizzuti2017An}. The FMMEM algorithm exhibits low sensitivity to parameter settings. The primary parameters, population size $N_{pop}$ and maximum number of iterations $t_{max}$, are empirically set at 100 and 500, respectively, to ensure performance while minimizing unnecessary computations. The parameters for the five algorithms under comparison are set according to the recommendations in \cite{Benson2016Higher, ma2019local, Li2021MotifbasedEL, li2019EdMot, Pizzuti2017An}. Specifically, the number of communities required by ME+k-means on each network is settled through trial-and-error experiments. In MELPA, the balance parameter $\alpha$ is set in range $[0, 1]$. EdMot incorporates the Louvain algorithm and sets the number of the largest connected components $k$ within the range $[1, 3]$. In MotifGA, the mutation rate is set at 0.1 and the crossover fraction at 0.8, while $N_{pop}$ and $t_{max}$ match those used in FMMEM.

\subsubsection{Real-world Networks}
The algorithms are tested on 9 typical real-world networks with different types, scales, and characteristics, of which more than 78\% of the networks have more than 1,000 edges. The PGP and Gplus networks have the largest scale, where PGP has 10,680 nodes and 24,316 edges, and Gplus has 23,628 nodes and 39,194 edges. Table \ref{table1} lists the detailed information of the 9 real-world networks, which can be consulted from the KONECT Project \cite{J2013KONECT} and the Network Data Repository \cite{Ryan2015The}.

\begin{table}[ht]
\setlength{\abovecaptionskip}{0.cm}
\setlength{\belowcaptionskip}{-0.cm}
\renewcommand\tabcolsep{14pt}
\renewcommand{\arraystretch}{1.00}
\caption{The Statistics of Real-world Networks Used in The Experiments. $n$ represents the number of nodes, $m$ represents the number of edges, $\left \langle d\right \rangle$ represents the average degree of nodes, and $cn$ is the number of communities.}
\centering
\begin{threeparttable}
\begin{tabular}{l c c c c}
\hline
\hline
Network       & $n$ & $m$ & $\left \langle d\right \rangle$ & $cn$\\
\hline
Karate 	          &34	      &78	      & 4.69            & 2    \\
Macaque              &47      &505         & 13.32           & 4    \\
Football 	         &115	       &613       & 10.66           & 12   \\
Email 	        & 1,133     &5,451       & 9.62           & -    \\
Polblogs           & 1,224	& 19,025	& 27.32        & 2    \\
Cora                  &2,708        &5,429      & 3.90            &7    \\
PowerGrid       &4,941         &6,594      & 2.67            & -   \\
PGP                 &10,680       &24,316	   & 4.55            & -   \\
Gplus	      &23,628	&39,194	& 3.32  & -    \\
\hline
\hline
\end{tabular}
\label{table1}
\begin{tablenotes}
       \footnotesize
       \item Note that '-' means that the ground-truth communities are not known.
     \end{tablenotes}
  \end{threeparttable}
\end{table}

\subsubsection{Synthetic Benchmark Networks}
The most widely used synthetic benchmark networks, i.e., Lancichinetti-Fortunato-Radicchi (LFR) \cite{Lancichinetti2008Benchmark}, are adopted for performance evaluation. In LFR networks, the number of nodes $n$ is set to 1,000. The mixing parameter $\mu$ varies from 0.1 to 0.9 with interval 0.1. The minimum and maximum community size (i.e., $c_{min}$ and $c_{max}$) are set to 20 and 50, and other parameters $\left \langle d\right \rangle$, $d_{max}$, $\tau_{1}$ and $\tau_{2}$ are fixed to 20, 50, 2 and 1, respectively. The ground truth partition of each LFR network can be automatically obtained according to predefined parameters during construction.

\subsubsection{Evaluation Metrics}
Two performance metrics, weighted modularity $Q_{W}$  and Normalized Mutual Information (NMI), are utilized to assess the quality of community partitions identified by different approaches. $Q_{W}$, as defined in Eq. \ref{equation9}, measures the quality of detected partitions without requiring ground-truth information \cite{Newman2004Analysis}. NMI assesses the accuracy of detected partitions by quantifying their similarity to the ground-truth partitions \cite{Lancichinetti2008Benchmark}, which can be formulated as
\begin{equation}
NMI(A,B) = \frac{-2\sum_{i = 1}^{c_A}\sum_{j = 1}^{c_B}N_{ij}{log\left ( \frac{N_{ij}n}{N_{i.}{N_{.j}}}\right)}}
{\sum_{i=1}^{c_A}N_{i.}{log \left ( \frac{N_{i.}}{n} \right)}+\sum_{j=1}^{c_B}N_{.j}log{ \left ( \frac{N_{.j}}{n}\right)}},
\label{equation13}
\end{equation}
where $c_{A}$ ($c_{B}$) denote the number of communities in partition $A$($B$), respectively. $N$ represents the confusion matrix, with each element $N_{i,j}$ indicating the number of nodes in the $i$th community of $A$ that appear in the $j$th community of $B$. $N_{i.}$ and $N_{.j}$ are the sum over the $i$th row and the $j$th column of matrix $N$, respectively. For both $Q_{W}$ and NMI, higher values indicate better detection outcomes.

\subsection{Performance on Synthetic Benchmark Networks}
\label{section5_2}

In this section, we assess the performance of FMMEM on synthetic benchmark networks, with the typical SOS serving as the baseline EA. The FMMEM’s performance is benchmarked against five state-of-the-art higher-order community detection algorithms: ME+k-means \cite{Benson2016Higher}, MLEO \cite{ma2019local}, MELPA \cite{Li2021MotifbasedEL}, EdMot \cite{li2019EdMot}, and MotifGA \cite{Pizzuti2017An}. A group of nine LFR networks with a scale of 1000 nodes are adopted, and the mixing parameter $\mu$ ranges from 0.1 to 0.9 with interval 0.1. On each LFR network, all algorithms employ the same triangle motif for fair comparison. 

\begin{figure}[ht]
\centering
\includegraphics[width=0.45\textwidth]{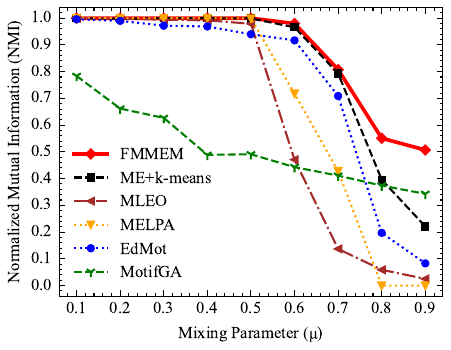}
\caption{The average NMI values derived from 20 independent runs of FMMEM and five state-of-the-art higher-order community detection algorithms over a group of LFR networks with $\mu$ increased from 0.1 to 0.9.}
\label{Fig5_FMMEM_LFR_NMI}
\end{figure}

Figure \ref{Fig5_FMMEM_LFR_NMI} presents the average NMI values derived from 20 independent runs of FMMEM and five other competitors on the LFR networks. FMMEM consistently achieves the highest NMI values, outperforming five competitors across all LFR networks. Specifically, when community structures are relatively distinct ($\mu$$\leq$$0.5$), most algorithms perform well, achieving NMI values close to 1. However, as the community structures become more ambiguous ($\mu$ increases from 0.6 to 0.9), all algorithms experience a decline in NMI values, while FMMEM maintains superior stability and accuracy. Particularly, between $0.8$$\leq$$\mu$$\leq$$0.9$, FMMEM’s NMI superiority over the competitors is significant, highlighting its effectiveness in managing ambiguous community structures. 

The empirical results demonstrate that FMMEM delivers competitive performance on synthetic LFR benchmarks, significantly outperforming several state-of-the-art higher-order community detection algorithms in terms of accuracy and stability, particularly in ambiguous community structures.

\begin{table*}[ht]
\setlength{\abovecaptionskip}{0.cm}
\setlength{\belowcaptionskip}{-0.cm}
\renewcommand\tabcolsep{10.0pt}
\renewcommand{\arraystretch}{1.00}
\caption{Average Optimal Value And Standard Deviation of $Q_{W}$ Obtained By FMMEM And Five State-Of-The-Art higher-order Community Detection Algorithms Performing 20 Independent Runs On nine Real-World Networks. The Best Solutions Obtained By All Algorithms On Each network Is Shown In Bold.}
\centering
\begin{threeparttable}
\begin{tabular}{l c c c c c c}
\hline
\hline
\multirow{2}*{Network}    & ME+k-means \cite{Benson2016Higher}  & MLEO \cite{ma2019local}   & MELPA \cite{Li2021MotifbasedEL}  & EdMot \cite{li2019EdMot}   & MotifGA \cite{Pizzuti2017An}  & FMMEM    \\
              ~                  & Mean(Std.)      & Mean(Std.)    & Mean(Std.)      & Mean(Std.)         & Mean(Std.)      & Mean(Std.) \\
\hline
Karate	& \bf{0.484(0.00e-00)}	&0.456(0.00e-00)	&0.458(0.00e-00)	&0.462(2.60e-02)	&0.465(9.28e-03)	& \bf{0.484(0.00e+00)}	\\
Macaque	&0.244(0.00e-00)		&0.000(0.00e-00)	&0.000(0.00e-00)	&0.257(1.18e-03)	&0.212(1.32e-02)	& \bf{ 0.265(0.00e+00)}	\\
Football	&0.852(0.00e-00)		&0.816(9.13e-03)	&0.833(1.58e-02)	&0.846(4.61e-03)	&0.827(1.42e-02)	& \bf{0.853(0.00e+00)}	\\
Email	&0.677(1.73e-03)		&0.018(3.47e-02)	&0.613(6.86e-02)	&0.675(9.47e-03)	&0.158(6.77e-02)	& \bf{0.701(5.00e -04)}	\\
Polblogs	&0.447(4.69e-03)		&0.445(4.66e-06)	& \bf{0.449(1.47e-05)}	&0.301(8.82e-02)	&0.328(3.16e-02)	& \bf{0.449(0.00e+00)}	\\
Cora		&0.863(2.15e-04)		&0.850(1.40e-03)	&0.863(8.73e-03)	&0.894 (2.65e-03)	&0.548(1.47e-02)	& \bf{0.923(2.34e-04)}	\\
PowerGrid	&0.925(4.21e-03)		&0.870(1.72e-03)	&0.895(8.06e-03)	&0.907(3.39e-03)	&0.721(8.77e-03)	& \bf{0.944(3.96e-04)}	\\
PGP		&0.721(1.35e-02)		&0.731(3.08-05)	&0.736(1.49e-02)	&0.797(4.38e-03)	&0.629(1.36e-02)	& \bf{0.809(2.44e-04)}	\\
Gplus	&0.418(1.39e-02)		&0.362(0.00e-00)	&0.408(2.28e-01)	&0.503(3.69e-03)	&0.236(1.34e-02)	& \bf{0.584(8.93e-05)}	\\
\hline
$p$-value  	&1.17E-02				& 7.69E-03		& 1.17E-02		&7.69E-03			&7.69E-03			&   -					\\
\hline
\hline
\end{tabular}
\label{table3}
\begin{tablenotes}
       \footnotesize
       \item Note that zero values indicate that all motifs are grouped into a single community, resulting in no effective motif-based partitions.
     \end{tablenotes}
  \end{threeparttable}
\end{table*}

\subsection{Performance on Real-World Networks}
\label{section5_3}

In this section, we further assess the performance of FMMEM on real-world networks, using the standard SOS serving as the baseline EA. The performance of FMMEM is compared with five state-of-the-art higher-order community detection algorithms: ME+k-means \cite{Benson2016Higher}, MLEO \cite{ma2019local}, MELPA \cite{Li2021MotifbasedEL}, EdMot \cite{li2019EdMot}, and MotifGA \cite{Pizzuti2017An}. Nine typical real-world networks with different types, ranging up to 23,628 nodes and 39,194 edges, are analyzed, with details provided in Table \ref{table1}. For each network, all algorithms employ the same triangle motif for fair comparison.

Table \ref{table3} records the average value and standard deviation of $Q_{W}$ for FMMEM and five competitors, based on 20 independent trials on each real-world network. FMMEM consistently yields the highest $Q_{W}$ values, outperforming five competitors in all nine real-world networks. Specifically, in four larger-scale networks (Cora, PowerGrid, PGP, and Gplus) with more than 2000 nodes, FMMEM’s average $Q_{W}$ improvement over the top competitor results is 5.73\%, exceeding the 1.36\% improvement observed in smaller networks. Particularly, in the largest Gplus network with 23,628 nodes and 39,194 edges, FMMEM’s average $Q_{W}$ improvement achieves 16.10\%.

The results in Table \ref{table3} are further evaluated using a typical nonparametric statistical procedure, i.e., Wilcoxon singed-rank test \cite{Biswas2017Analyzing, Biswas2016A}, for checking the statistical difference between FFMEM and five competitors. Statistical analysis shows that, at a significance level of 0.05, FFMEM significantly outperforms the five competitors. The findings prove the effectiveness of FFMEM in discovering high-quality motif-based community partitions in real-world networks, and the advantages in relatively larger-scale networks. 

\subsection{Ablation Analysis}
\label{section5_4}
The superior performance of FMMEM has been demonstrated through the experimental results presented above. To further assess the effect of two key strategies in FMMEM, namely FM-NCM and FM-LCM, an ablation analysis is conducted. Specifically, three algorithm frameworks are developed to apply different combinations of strategies: (i) Basic: utilizes the original EAs for motif-based modularity optimization; (ii) Basic+NCM: incorporates the FM-NCM strategy to correct misassigned bridge nodes and enhance higher-order partition quality; and (iii) FMMEM (Basic+NCM+LCM): our proposed framework, which additionally integrates the FM-LCM strategy to merge excessively fragmented communities and improve local community search. All the three frameworks employ four baseline EAs: genetic algorithm (GA), differential evolution (DE), farmland fertility (FF), and symbiotic organisms search (SOS), enabling the evaluation of the generality of FM-NCM and FM-LCM strategies.

\begin{figure}[ht]
\centering
\includegraphics[width=0.50\textwidth]{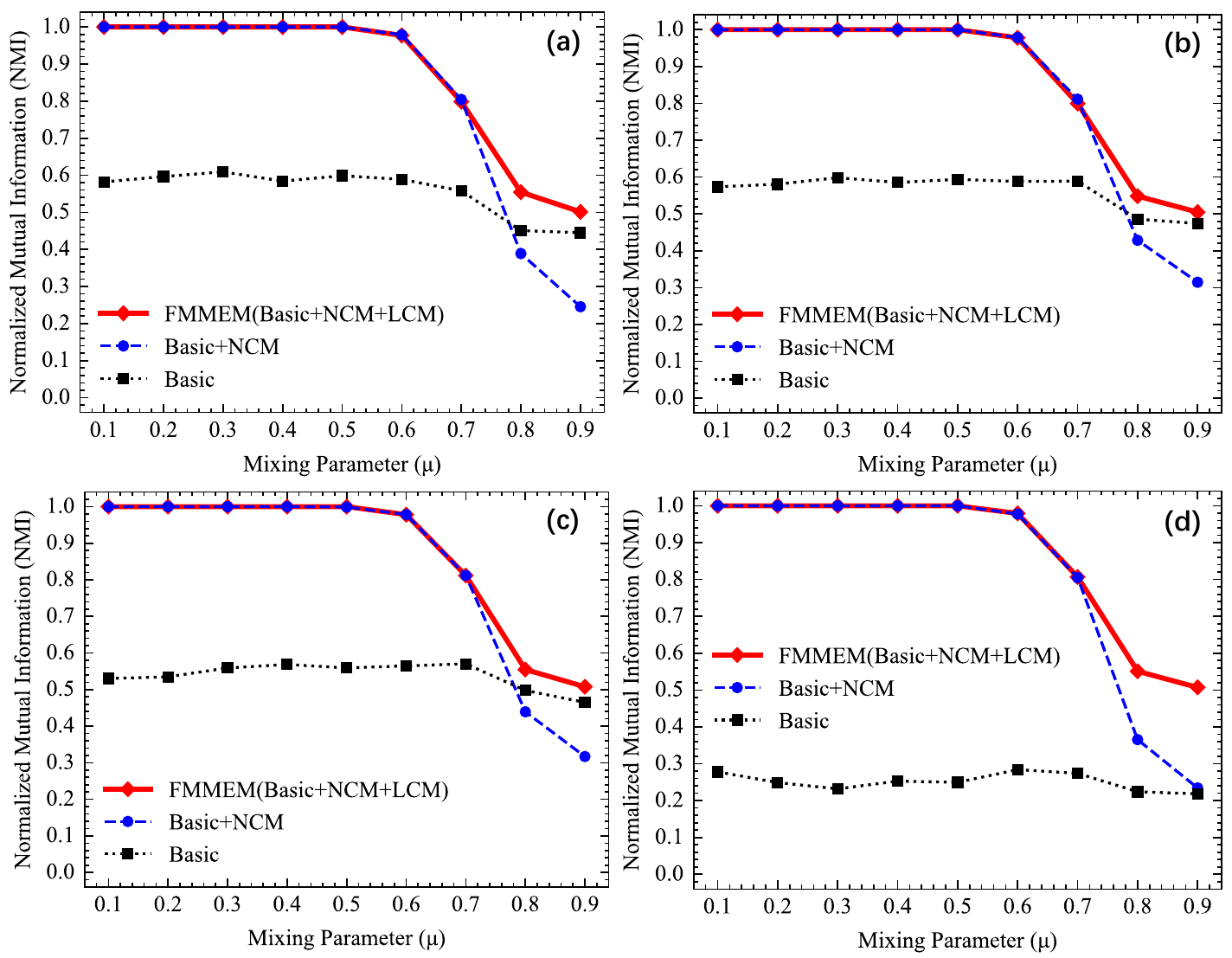}
\caption{The average NMI values obtained by FMMEM and two compared frameworks (i.e., Basic and Basic+NCM) performing 20 independent runs based on four baseline EAs on a group of LFR networks with $\mu$ values ranging from 0.1 to 0.9. (a) GA, (b) DE,  (c) FF, and (d) SOS.}
\label{Fig6_FMMEM_NCM_LCM_LFR}
\end{figure}

First, the ablation analysis of the proposed FMMEM (Basic+NCM+LCM) framework is validated across a group of LFR networks with $\mu$ values increasing from 0.1 to 0.9. For fair comparison, FMMEM and two compared frameworks (i.e., Basic and Basic+NCM), employ the same triangle motif $M\left (3,3\right )$. The average NMI values, obtained by each framework performing 20 independent runs based on four baseline EAs on each LFR network, are illustrated in Fig. \ref{Fig6_FMMEM_NCM_LCM_LFR}. 

Figure \ref{Fig6_FMMEM_NCM_LCM_LFR} shows that FMMEM consistently outperforms the other two frameworks in all test cases. Specifically, when employing a particular baseline EA, both FMMEM and Basic+NCM exhibit significant advantages over Basic in NMI values when $\mu$ ranges from 0.1 to 0.7, indicating obvious accuracy improvement. moreover, FMMEM obtains higher NMI values than Basic+NCM when $\mu$ increasing from 0.7 to 0.9. The findings validate the effectiveness of the FM-NCM and FM-LCM strategies in enhancing accuracy of FMMEM. Particularly, FM-NCM shows stronger effect in enhancing overall higher-order partition accuracy, while FM-LCM further promotes approximation to the optimum in relatively ambiguous structures. Additionally, the consistent accuracy superiority of FMMEM observed with various baseline EAs prove the generality of the FM-NCM and FM-LCM strategies.

\begin{table}[ht]
\setlength{\abovecaptionskip}{0.cm}
\setlength{\belowcaptionskip}{-0.cm}
\renewcommand\tabcolsep{1.0 pt}
\renewcommand{\arraystretch}{1.00}
\caption{Average Optimal Value And Standard Deviation of $Q_{W}$ Obtained By FMMEM And Two Compared Frameworks (i.e., Basic and Basic+NCM) Performing 20 Independent Runs Based On Four Baseline EAs. The Best Solutions Obtained By All Three Algorithm Frameworks With Each Baseline EA On Each network Is Shown In Bold.}
\centering
\begin{tabular}{l c c c c}
\hline
\hline
\multirow{2}*{Network}    & \multirow{2}*{Algorithm}    & Basic         	        & Basic+NCM        	& FMMEM \\
                        ~                   &   ~                                            & Mean(Std.)     &  Mean(Std.)      & Mean(Std.) \\
\hline

\multirow{4}*{Karate}       &GA		&0.466(8.78e-03)		&\bf{0.484(0.00e+00)}	&\bf{0.484(0.00e+00)} \\
					&DE		&0.427(3.69e-03)		&\bf{0.484(0.00e+00)}	&\bf{0.484(0.00e+00)} \\
					&FF		&\bf{0.484(0.00e+00)}	&\bf{0.484(0.00e+00)}	&\bf{0.484(0.00e+00)} \\
					&SOS	&0.452(1.49e-02) 		&\bf{0.484(0.00e+00)}       &\bf{0.484(0.00e+00)} \\
					
\hline
\multirow{4}*{Macaque}   &GA		&0.255(2.46e-03)	&\bf{0.265(0.00e+00)}	&\bf{0.265(0.00e+00)} \\
					&DE		&0.133(2.16e-02)	&\bf{0.265(0.00e+00)}	&\bf{0.265(0.00e+00)} \\
					&FF		&0.233(3.02e-02)	&\bf{0.265(0.00e+00)}	&\bf{0.265(0.00e+00)} \\
					&SOS	&0.251(4.41e-03)	&\bf{0.265(0.00e+00)}       &\bf{0.265(0.00e+00)} \\
\hline
\multirow{4}*{Football}	&GA  	&0.295(3.65e-02)	&\bf{0.853(0.00e+00)}	&\bf{0.853(0.00e+00)} \\					
					&DE	 	&0.168(9.60e-03)	&\bf{0.853(0.00e+00)}	&\bf{0.853(0.00e+00)} \\
					&FF		&0.131(1.17e-02)	&\bf{0.853(0.00e+00)}	&\bf{0.853(0.00e+00)} \\
					&SOS	&0.573(7.37e-02)	&\bf{0.853(0.00e+00)}	&\bf{0.853(0.00e+00)} \\
					
\hline
\multirow{4}*{Email}	&GA  	&0.151(2.17e-02)	&0.687(3.22e-03)	&\bf{0.701(4.83e-04)}\\			
					&DE	 	&0.063(1.28e-02)	&0.685(1.70e-03)	&\bf{0.701(5.45e-04)} \\
					&FF		&0.060(1.25e-02)	&0.687(3.98e-03)	&\bf{0.701(2.69e-04)} \\
					&SOS	&0.324(6.87e-02)	&0.686(3.16e-03)	&\bf{0.702(1.77e-04)} \\

\hline
\multirow{4}*{Polblogs}	&GA  	&0.180(3.12e-02)	&\bf{0.449(0.00e+00)}	&\bf{0.449(0.00e+00)}	\\					
					&DE	 	&0.119(2.88e-02)	&\bf{0.449(0.00e+00)}	&\bf{0.449(0.00e+00)}	 \\
					&FF		&0.105(2.59e-02)	&\bf{0.449(0.00e+00)}	&\bf{0.449(0.00e+00)}	 \\
					&SOS	&0.291(2.91e-02)	&\bf{0.449(0.00e+00)}	&\bf{0.449(0.00e+00)}	 \\
					
\hline
\multirow{4}*{Cora}		&GA  	&0.079(9.03e-03)	&0.880(2.32e-03)	&\bf{0.919(4.76e-04)}\\					
					&DE	 	&0.022(2.45e-02)	&0.881(1.74e-03)	&\bf{0.921(5.00e-04)}	 \\
					&FF		&0.021(2.51e-02)	&0.882(1.91e-03)	&\bf{0.922(5.03e-04)}\\
					&SOS	&0.076(3.49e-02)	&0.880(2.63e-03)	&\bf{0.923(5.19e-04)}\\

\hline
\multirow{4}*{PowerGrid}		&GA  	&0.164(1.37e-02)	&0.889(3.95e-03)	&\bf{0.941(1.09e-03)}\\					
						&DE	 	&0.027(1.04e-02)	&0.891(4.77e-03)	&\bf{0.943(7.12e-04)}\\
						&FF		&0.020(1.06e-02)	&0.893(5.06e-03)	&\bf{0.943(1.05e-03)}\\
						&SOS	&0.109(3.08e-02)	&0.889(2.87e-03)	&\bf{0.944(3.96e-04)}\\

\hline
\multirow{4}*{PGP}			&GA  	&0.198(4.90e-02)	&0.791(4.96e-04)	&\bf{0.808(1.17e-04)}\\					
						&DE	 	&0.089(4.28e-02)	&0.790(5.27e-04)	&\bf{0.807(3.96e-04)}\\
						&FF		&0.087(3.79e-02)	&0.791(4.31e-04)	&\bf{0.808(1.23e-04)}\\
						&SOS	&0.164(3.92e-02)	&0.790(9.07e-04)	&\bf{0.809(2.44e-04)}\\

\hline
\multirow{4}*{Gplus}		&GA  	&0.109(2.82e-03)	&0.577(1.87e-03)	&\bf{0.584(3.29e-04)}\\					
						&DE	 	&0.029(1.69e-02)	&0.577(1.44e-03)	&\bf{0.584(1.59e-04)}\\
						&FF		&0.012(7.93e-03)	&0.578(2.58e-03)	&\bf{0.584(1.04e-04)}\\
						&SOS	&0.109(7.82e-03)	&0.577(7.31e-04)	&\bf{0.584(8.93e-05)}\\
\hline
\multicolumn{2}{c}{$p$-value}		& 2.47E-07	& 8.70E-05		&   -				\\
\hline
\hline
\end{tabular}
\label{table2}
\end{table}

Second, the ablation analysis of the proposed FMMEM (Basic+NCM+LCM) framework is further validated on real-world networks. Table \ref{table2} records the average optimal value and standard deviation of $Q_{W}$ obtained by each framework performing 20 independent runs based on four baseline EAs on each real-world network. The results demonstrates that FMMEM consistently achieves the highest $Q_{W}$ values compared to the other two frameworks in the majority of test cases. Specifically, FMMEM significantly outperforms Basic in $Q_{W}$ values across all nine real networks and exceeds Basic+NCM in five (55.6\%) networks. The $p$-value of the Wilcoxon Signed-Rank test is always less than 0.05, further confirming the quality superiority of FMMEM compared to Basic and Basic+NCM. Particularly, in the Email network and the four relatively larger-scale networks, Cora, PowerGrid, PGP, and Gplus, the improvement of FMMEM in $Q_{W}$ values is more significant. Furthermore, the consistent quality improvements of FMMEM observed across various baseline EAs confirm the universal applicability of the FM-NCM and FM-LCM strategies.

The results on both synthetic and real-world networks indicate that the FM-NCM and FM-LCM strategies in FMMEM are truly effective in enhancing the accuracy and quality of higher-order partitions optimized by EAs. Consequently, these strategies enable FMMEM to achieve superior higher-order partitions, particularly in ambiguous community structures. The superiority does not depend on the choice of the baseline EA in FMMEM. The success of FM-NCM and FM-LCM primarily stems from the effective use of higher-order fuzzy community information, applicable to other higher-order community detection algorithms as well.

\section{Conclusion}
\label{section6}

In summary, we study the challenges of higher-order community detection in improving the quality of identified higher-order communities and providing more rich and fine-grained higher-order fuzzy community information. To address these issues, we first introduce a new concept of higher-order fuzzy memberships, which quantify the membership grades of motifs to crisp higher-order communities. The higher-order fuzzy memberships accurately reflect partial topological or functional belonging of each single motif to multiple communities, thereby facilitating an understanding of fuzzy characteristics of higher-order community structures. Furthermore, we employ these higher-order fuzzy memberships as more detailed assistant information to enhance HCD via a general framework called fuzzy memberships assisted motif-based evolutionary modularity (FMMEM). Compared to state-of-the-art HCD methods, FMMEM significantly improves the quality of the identified higher-order communities, particularly in networks with ambiguous structures and relatively large scales.

First, higher-order fuzzy memberships are defined for quantifying membership grades within interval $\left [0,1 \right ]$ of motifs to crisp higher-order communities, providing richer and more fine-grained higher-order fuzzy community information. Second, higher-order fuzzy memberships are employed in FMMEM as detailed assistant information to improve quality of optimized higher-order partitions. A fuzzy memberships-based neighbor community modification (FM-NCM) strategy is designed to correct misassigning bridge nodes for improving partition quality. At last, a fuzzy memberships-based local community merging (FM-LCM) strategy is developed to merge excessively fragmented communities, enhancing local search to approximate the global optimum. Experimental results indicate the effectiveness of higher-order fuzzy memberships in reflecting partial community affiliations of motifs, and the superiority of FMMEM in identifying high-quality higher-order communities in both synthetic and real-world datasets.

This work highlights the significance of higher-order fuzzy memberships in enriching crisp higher-order community structures, as well as the potential of leveraging higher-order fuzzy memberships to improve HCD performance. The fine-grained higher-order fuzzy community information is applicable to other research issues, facilitating more precise network topology and functional analysis. Additionally, high-quality higher-order partitions can provide more realistic community information for deeper network applications, such as motif-based link prediction \cite{Li2022LP}, brain network analysis \cite{Duclos2021Brain}, and EEG-based human intention recognition \cite{Zhang2020MakingSO, Luo2018AnAS,Chen2020ASR}, etc. In future research, we will focus on HCD in larger-scale real-world networks with complex and ambiguous structures, and further study higher-order fuzzy overlapping community detection.

\balance
\bibliography{FMMEM_references}

\end{document}